\documentclass[english]{cccconf}
\usepackage[comma,numbers,square,sort&compress]{natbib}
\usepackage{epstopdf}

\begin{document}
	
	\title{Adaptive Parameter Control Using AAN for Lower Limb Rehabilitation Exoskeletons}
	
	\author{%
		Zheng Sun\aref{amss},
		Wenkong Wang\aref{amss},
		Zizhong Wei*\aref{otherinst},
		Xin Ma*\aref{amss}\thanks{Corresponding author: maxin@sdu.edu.cn}
	}
	
	\affiliation[amss]{%
		Center for Robotics, School of Control Science and Engineering,\\
		Engineering Research Center of Intelligent Unmanned System, Ministry of Education,\\
		Shandong University,\\
		No.17923, Jingshi Road(s), Jinan, 250061, China
	}
	
	\affiliation[otherinst]{%
		Shandong Inspur Science Research Institute Co., Ltd.
	}
	
	\maketitle
	
\begin{abstract}
	Exoskeletons play a crucial role in assisting patients with varying mobility levels during rehabilitation. However, existing control strategies face challenges such as imprecise trajectory tracking, interaction torque oscillations, and limited adaptability to diverse patient conditions. To address these issues, this paper proposes an assist-as-needed (AAN) control algorithm that integrates a human-robot coupling dynamics model, a human torque-momentum observer (HTMO), and an adaptive parameter controller (APC). The algorithm first employs inverse dynamics to compute the joint torques required for the rehabilitation trajectory. The HTMO then estimates the torque exerted by the patient’s joints and determines the torque error, which the exoskeleton compensates for via a spring-damper system, ultimately generating the target trajectory. Finally, the APC ensures adaptive assistive control. The proposed method is validated for its effectiveness in Matlab/Simulink.
\end{abstract}
	
	
	\footnotetext{This work was supported in part by the National Key Research and Development Program (No. 2023YFB4706104).}

\section{Introduction}
Stroke is one of the leading causes of disability and death worldwide, with its incidence rising annually, especially in aging societies[1]. After a stroke, patients often experience hemiplegia and muscle weakness, which severely affect their quality of life. Traditional rehabilitation methods, primarily physical therapy, aid recovery but are limited by treatment intensity and duration. With the growing shortage of medical resources, exoskeleton-assisted rehabilitation has emerged as an innovative technology, demonstrating significant success in providing personalized and efficient rehabilitation support while helping patients recover motor function[2][3].

Exoskeleton-assisted rehabilitation therapy is categorized into passive and active modes. In the passive mode[4], the exoskeleton assists the patient in performing rehabilitation exercises without requiring active exertion, typically employed in the early stages of rehabilitation. In the active mode[5], the exoskeleton provides assistance through a power system, enabling the patient to actively exert force, thereby facilitating the recovery of muscle strength and motor coordination. By incorporating the AAN strategy, the exoskeleton can dynamically adjust the level of assistance based on the patient's real-time needs, offering precise support and avoiding both excessive and insufficient assistance, thus optimizing rehabilitation outcomes[6][7]. When evaluating the patient's movement ability, trajectory tracking error can serve as an indicator of recovery capacity; the smaller the error, the better the patient's movement ability[8].

In recent years, active control, particularly AAN control, has become a research hotspot in academia. Numerous studies have focused on acquiring human motion intentions using various approaches, such as force/torque sensors[9][10], electroencephalography (EEG)/electromyography (EMG) signals[11][12], and inertial measurement unit (IMU) sensors. Although these methods show some potential, they still face some limitations in practical applications. For instance, force/torque sensors are expensive, require precise installation, and can constrain human-machine interactions. Similarly, EEG and EMG signals, though informative, suffer from low signal-to-noise ratios and susceptibility to interference, limiting their reliability in real-world settings.

To overcome these challenges, recent studies[7][13] have explored alternative strategies, such as motion intention observation based on interactive torque observers. These methods use techniques like Kalman filtering to mitigate modeling inaccuracies and improve estimation reliability. Despite these advances, the main problems with existing methods are as follows: 1) Many solutions rely on external sensors, which introduce complexity and increase system cost. 2)Existing approaches lack sufficient adaptability to accommodate individual user differences, leading to suboptimal performance across diverse users.

In comparison to the previous work this paper makes the following improvements:

1) A sensorless joint torque estimation method is introduced, which leverages system dynamics and real-time feedback, thereby eliminating the reliance on external sensors. This reduction in hardware complexity and cost is achieved without compromising the accuracy of torque estimation.

2) An adaptive control strategy is developed to address inter-user variability, ensuring robust performance across individuals with varying biomechanical characteristics. This enhances the system's versatility and reliability, thereby expanding its applicability to a diverse user population.

The structure of this paper is as follows: Section 2 describes the Human-machine coupled dynamics systems; In Section 3, the torque observer and the target controller are given; Simulation results and discussion  are provided in Section 4; Section 5 provides a summary.

\section{Problem Formulation}
\subsection{Human/Exoskeleton model}
The model in Fig. 1 (left) illustrates the rehabilitation exoskeleton, consisting of hip and knee joint modules and support straps. It assists hip and knee flexion/extension during gait rehabilitation, with control based on lower limb dynamics[15], as detailed below:
\begin{eqnarray}
&{M_i}({q_i}){{\ddot q}_i} + {C_i}({{\dot q}_i},{q_i}){{\dot q}_i} + {G_i}({q_i}) = {\tau _i}{\rm{  }} \quad i = h \, or \, e
\end{eqnarray}
where $M_i(q_i) $ is the joint inertia matrix, $ C_i(\dot{q}_i, q_i) $ represents coriolis matrix, $ G_i(q_i) $ is the gravity matrix, and $ \tau_i $ is the control input torque. $ q_i $, $ \dot{q}_i $, and $ \ddot{q}_i $ denote joint angle, angular velocity, and acceleration, respectively. $ i $ can be $ h $ (human) or $ e $ (exoskeleton), distinguishing their kinetic parameters.
The expressions are given as:  
\begin{eqnarray}
{M_i}& =& \left[ \begin{array}{l}
	{m_{i11}}\quad{m_{i12}}\\\nonumber
	{m_{i21}}\quad{m_{i22}}
\end{array} \right],{C_i} = \left[ \begin{array}{l}
	{c_{i11}}\quad{c_{i12}}\\
	{c_{i21}}\quad\,\,0
\end{array} \right],\\
{G_i} &=& \left[ {{g_{i1}}\quad{g_{i2}}} \right]^T
\end{eqnarray}
where ${m_{i11}}={m_{i2}}L_{i1}^2 + 2{m_{i2}}{C_{i2}}{L_{i1}}{l_{i2}} + {m_{i1}}l_{i1}^2 + {m_{i2}}l_{i2}^2 + {J_{i1}} + {J_{i2}}$, ${m_{i12}}={m_{i21}}= {m_{i2}}l_{i2}^2 + {L_{i1}}{m_{i2}}{C_{i2}}{l_{i2}} + {J_{i2}}$, ${m_{i22}}={m_{i2}}l_{i2}^2 + {J_{i2}}$, ${c_{i11}}= - 2{L_{i1}}{l_{i2}}{m_{i2}}{S_{i2}}{{\dot \theta }_{i2}}$, ${c_{i12}}=-{L_{i1}}{l_{i2}}{m_{i2}}{S_{i2}}{{\dot \theta }_{i2}}$, ${L_{i1}}{l_{i2}}{m_{i2}}{S_{i2}}{{\dot \theta }_{i1}}$, 
${G_{i1}}=-g{l_{i2}}{m_{i2}}{C_{i1i2}} - {L_{i1}}g{m_{i2}}{C_{i1}} - g{l_{i1}}{m_{i1}}{C_{i1}}$,
${G_{i2}}={ - g{l_{i2}}{m_{i2}}{C_{i1i2}}}$. The variables \( q_i = [\theta_{i1}, \theta_{i2}]^T \), \( \dot{q}_i = [\dot{\theta}_{i1}, \dot{\theta}_{i2}]^T \), and \( \ddot{q}_i = [\ddot{\theta}_{i1}, \ddot{\theta}_{i2}]^T \). The trigonometric terms like $ S_{ij} $ and $ C_{ij} $ denote $\sin(\theta_{ij})(j=1,2)$ and $\cos(\theta_{ij})(j=1,2)$, respectively. The complete notation definitions are given in \textbf{Table1}.

\begin{figure}[!htb]
	\centering
	\includegraphics[width=\hsize]{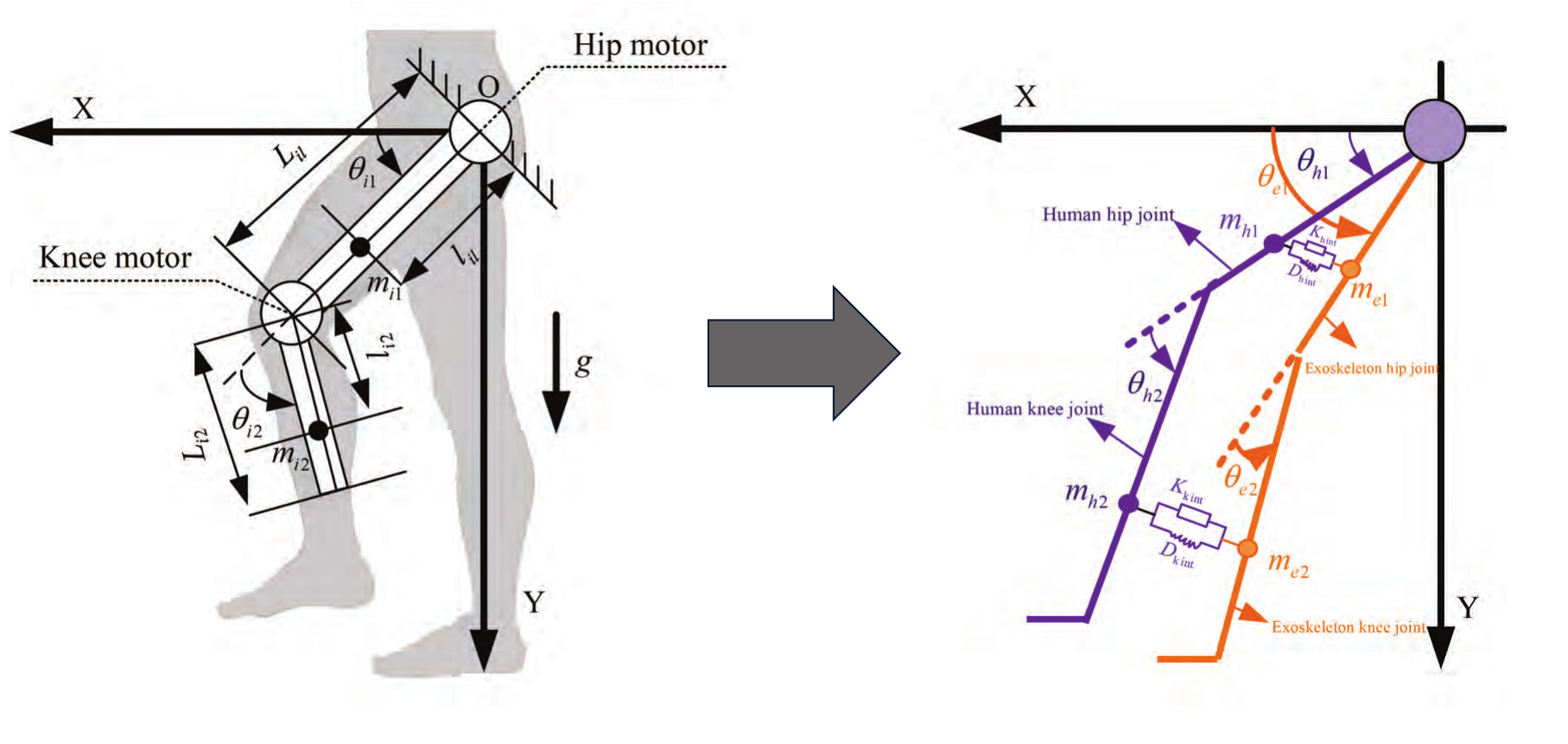}
	\caption{Human-machine coupled dynamics system}
	\label{figure1}
\end{figure}

\subsection{Coupling model of interaction forces}

As shown in Fig. 1 (right), the exoskeleton-human coupling is modeled using a spring-damping system, assuming aligned hip/knee joints connected by straps. The interaction torque is derived as follows[15]:
\begin{eqnarray}
	{F_h} 
 = {K_{{\rm{h}}{\mathop{\rm int}} }}\left[ \begin{array}{l}
		{x_{e1}} - {x_{h1}}\\
		{y_{e1}} - {y_{e1}}
	\end{array} \right] + {D_{{\rm{h}}{\mathop{\rm int}} }}\left[ \begin{array}{l}
		{{\dot x}_{e1}} - {{\dot x}_{h1}}\\
		{{\dot y}_{e1}} - {{\dot y}_{h1}}
	\end{array} \right]
\end{eqnarray}
\begin{eqnarray}
	{F_k} = {K_{{\rm{k}}{\mathop{\rm int}} }}\left[ \begin{array}{l}
		{x_{e2}} - {x_{h2}}\\
		{y_{e2}} - {y_{h2}}
	\end{array} \right] + {D_{{\rm{k}}{\mathop{\rm int}} }}\left[ \begin{array}{l}
		{{\dot x}_{e2}} - {{\dot x}_{h2}}\\
		{{\dot y}_{e2}} - {{\dot y}_{h2}}
	\end{array} \right]
\end{eqnarray}

Where $F_h$ and $F_k$ represent the interaction forces at the hip and knee joints, respectively. $K_{{\rm hint}}$ and $ K_{{\rm kint}}$ are the stiffness parameters, while $D_{{\rm hint}} $ and $D_{{\rm kint}}$ are the damping parameters for the hip and knee joints. The coordinates of the centers of mass for the human thigh and shank, as well as the exoskeleton’s thigh and shank, are denoted as $H_h(x_{h1}, y_{h1})$, $H_k(x_{h2}, y_{h2})$, $E_h(x_{e1}, y_{e1})$, and $E_k(x_{e2}, y_{e2})$, respectively. The specific expressions are as follows:
\begin{eqnarray}\nonumber
&&	\left\{ \begin{array}{l}
		{x_{i1}} = {\rm{ }}{l_{i1}}{S_{i1}}\\
		{y_{i1}} = {l_{i1}}{C_{i1}}\\
		{x_{i2}} = {L_{i1}}{S_{i1}} + {l_{i2}}{S_{i1i2}}\\
		{y_{i2}} = {L_{i1}}{C_{i1}} + {l_{i2}}{C_{i1i2}}
	\end{array} \right.\\
\Rightarrow &&\left\{ \begin{array}{l}
		{{\dot x}_{i1}} = {\rm{ }}{l_{i1}}{C_{i1}}{{\dot \theta }_{i1}}\\
		{{\dot y}_{i1}} =  - {l_{i1}}{S_{i1}}{{\dot \theta }_{i1}}\\
		{{\dot x}_{i2}} = {L_{i1}}{C_{i1}}{{\dot \theta }_{i1}} + {l_{i2}}{C_{i1i2}}({{\dot \theta }_{i1}} + {{\dot \theta }_{i2}})\\
		{{\dot y}_{i2}} =  - {L_{i1}}{S_{i1}}{{\dot \theta }_{i1}} - {l_{i2}}{S_{i1i2}}({{\dot \theta }_{i1}} + {{\dot \theta }_{i2}})
	\end{array} \right.
\end{eqnarray}
Next, the contact forces at the centers of mass, defined by Eqs. (3) and (4), are then converted to joint torques using the Jacobian matrix. The resulting expression is as follows:
\begin{eqnarray}
	{\tau _{Jh}} &=& J_{hh}^T{F_h} + J_{kh}^T{F_k}\\
	{\tau _{Je}} &=& - J_{he}^T{F_h} - J_{ke}^T{F_k}
\end{eqnarray}
where $J_{hi}$ and $J_{ki}$ denote the Jacobian matrices of the contact points.

\begin{table}[t]
	\begin{center}
		\renewcommand{\arraystretch}{1.2}
		\caption{System parameters}
		\label{tab_3}
		\begin{tabular}{c c}
			\hline
			\normalsize \textbf{Parameters}  &\normalsize \textbf{Physical meaning} 	 \\
			\hline
			
			\normalsize${L_{i1}}$  &\normalsize Length of thigh   \\
			
			\normalsize$\l_{i1}$ &\normalsize Length from hip to thigh center of mass \\
			
			\normalsize $m_{i1}$  & \normalsize Mass of thigh  \\ 
			
			\normalsize $J_{i1}$  & \normalsize Hip joint moment of inertia   \\ 
			
			\normalsize$\theta_{i1}$ & \normalsize Hip rotation angle \\
			
			\normalsize${L_{i2}}$ & \normalsize Length of shank   	 \\
			
			\normalsize$\l_{i1}$ & \normalsize Length from knee to thigh center of mass   \\
			
			\normalsize $m_{i2}$  & \normalsize Mass of shake  	   \\
			
			\normalsize$\theta_{i2}$ & \normalsize Knee rotation angle  \\
			
			\normalsize $J_{i2}$  & \normalsize Knee joint moment of inertia   \\ 
			
			\normalsize$g$ & \normalsize Gravitational acceleration \\
			
			\normalsize$K$ & \normalsize System coupling stiffness \\
			
			\normalsize$D$ & \normalsize System coupling damping \\
			
			\hline
		\end{tabular}
	\end{center}
\end{table}

\section{Main Results}
\subsection{HTMO Design}
Considering the uncertainties in human system modeling and the nonlinear, coupled effects in human-computer interaction, Eq. (1) of the human wear dynamics model can be revised as:
\begin{eqnarray}
{M_d}({q_h}){{\ddot q}_h} + {C_d}({{\dot q}_h},{q_h}){{\dot q}_h} + {G_d}({q_h}) = {\tau_{htotal}}  + {\tau _{Jh}}
\end{eqnarray}
where ${M_d}({q_h}) = {M_h}({q_h}) + \Delta {M_h}({q_h})$, ${C_d}({{\dot q}_h},{q_h}) = {C_h}({{\dot q}_h},{q_h}) + \Delta {C_h}({{\dot q}_h},{q_h})$, ${G_d}({q_h}) = {G_h}({q_h}) + \Delta {G_h}({q_h})$, ${\tau _{htotal}} = {\tau _h} + {\tau _d}$, ${\tau _{htotald}} = {\tau _{htotal}} + {\tau_{Jhd}}$, ${\tau _d}$ is an unmodelled disturbance. $ M_d $ still satisfies the following properties[13]: 1) positive definite inertia matrix; 2) ${{\dot M}_d}(q) = {C_d}({q_h},{{\dot q}_h}) + C_d^T({q_h},{{\dot q}_h})$.

The definition of the human body's generalized momentum is as follows:
\begin{eqnarray}
P = {M_d}{\dot q_h}
\end{eqnarray}
Take the time derivative of $P$ :
\begin{eqnarray}\nonumber
\dot P &=& {{\dot M}_d}{{\dot q}_h} + {M_d}{{\ddot q}_h}\\
 &=& {\tau _{htotal}} + {\tau _{Jh}} - {G_d} + C_d^T{{\dot q}_h}
\end{eqnarray}
Inspired by [13]:
\begin{eqnarray}
	\dot {\hat P} &=&  {T_o} + {\tau _{Jh}}  - {G_d} + C_d^T{{\dot q}_h}\\
	{T_o} &=& \ell(P - \hat P)
\end{eqnarray}
where $T_o$ is called the residual vector.
From Eq. (9)-(12), it follows that:
\begin{eqnarray}
	{{\dot T}_o} = \ell({\tau _{htotal}} - {T_o})
\end{eqnarray}
Next, a simultaneous Laplace transform on both sides of Eq. (13) is obtained:
\begin{eqnarray}
{\tau _{htotal}}(s) = \frac{{s + {\ell}}}{{{\ell}}}{T_o}(s)
\end{eqnarray}
It can be seen that ${\ell} \to  + \infty $, ${T_o} \to {\tau _{htotal}}$.

\begin{figure}[t]
	\centering
	\includegraphics[width=\hsize]{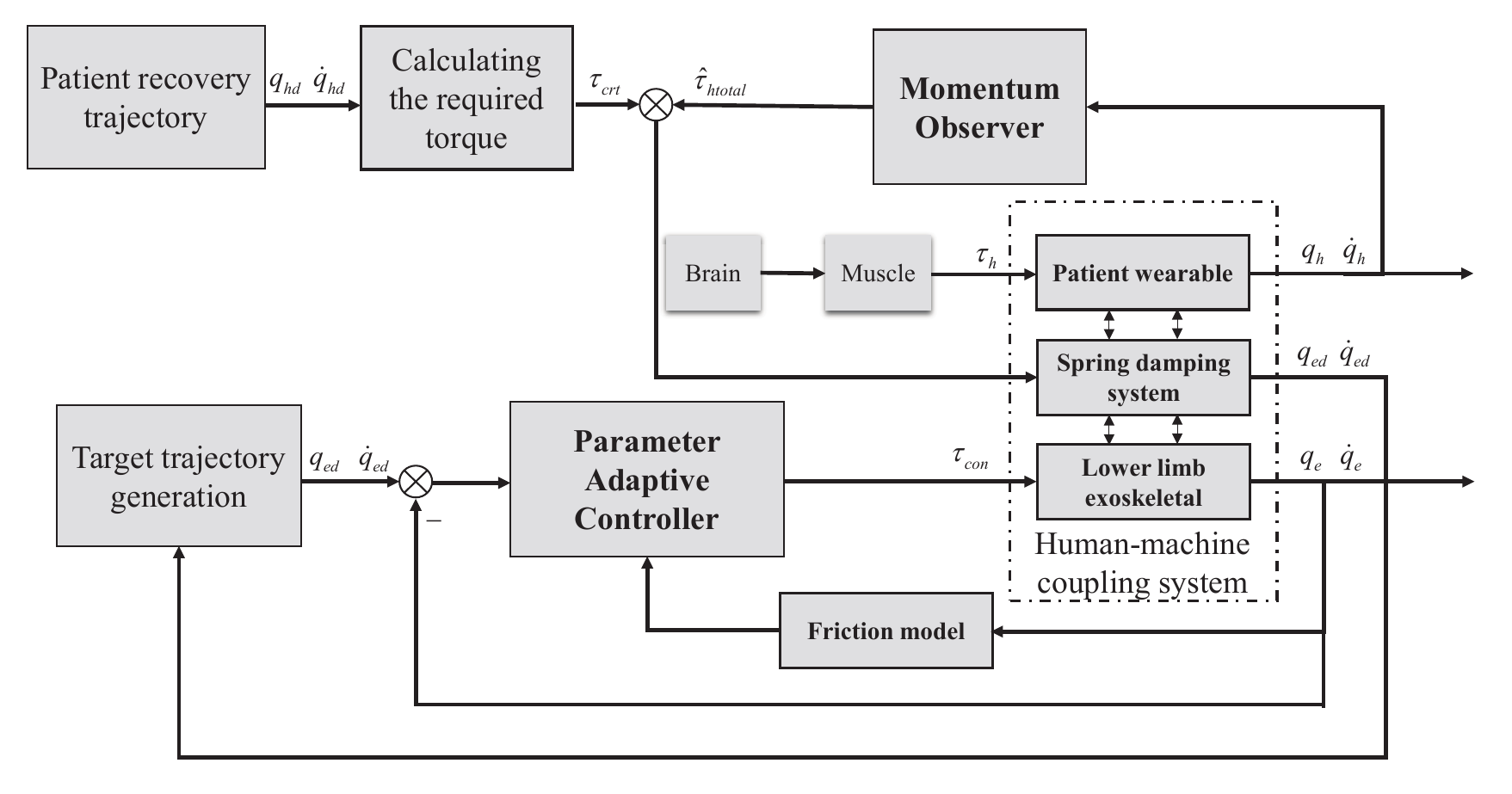}
	\caption{System control logic block diagram}
	\label{figure2}
\end{figure}
From Fig. 2, it can be concluded:
\begin{eqnarray}
	{\tau _{crt}} = {\hat \tau _{htotal}} + {\tau _{Jhd}}
\end{eqnarray}
where target interaction torque ${\tau _{Jhd}}$ is considered as assistance torque.  

In an ideal scenario, when the patient has full voluntary control over their movements, the joint torque generated by the human body should satisfy the condition ${{ \tau }_{htotal}} = {\tau _{crt}}$. However, most patients do not have full muscle control, i.e., ${{ \tau }_{htotal}} < {\tau _{crt}}$, and the greater the difference between ${{ \tau }_{htotal}}$ minus ${\tau _{crt}}$, the greater the required assistive torque. Therefore, during the rehabilitation process, assistive control strategies can be implemented by adjusting the desired interaction torque to provide effective support and facilitate the recovery of motor functions, thereby offering appropriate assistance to the patient.

Based on Eq. (15), the desired target interaction torque $ {\tau _{Jhd}}$ can be determined. Subsequently, the relationship between the interaction torque and the exoskeleton will be discussed. By deriving and transforming Eqs. (3) and (4) to (7), the following results can be obtained:
\begin{eqnarray}\nonumber
{\tau _{Jhd}} = J_{hh}^T({K_{{\rm{hint}}}}\left[ {\begin{array}{*{20}{l}}
		{{x_{e1d}} - {x_{h1}}}\\
		{{y_{e1d}} - {y_{e1}}}
\end{array}} \right] + {D_{{\rm{hint}}}}\left[ {\begin{array}{*{20}{l}}
		{{{\dot x}_{e1d}} - {{\dot x}_{h1}}}\\
		{{{\dot y}_{e1d}} - {{\dot y}_{h1}}}
\end{array}} \right])
\end{eqnarray}
\begin{eqnarray}
 + J_{hk}^T({K_{{\rm{kint}}}}\left[ {\begin{array}{*{20}{l}}
			{{x_{e2d}} - {x_{h2}}}\\
			{{y_{e2d}} - {y_{h2}}}
	\end{array}} \right] + {D_{{\rm{kint}}}}\left[ {\begin{array}{*{20}{l}}
			{{{\dot x}_{e2d}} - {{\dot x}_{h2}}}\\
			{{{\dot y}_{e2d}} - {{\dot y}_{h2}}}
	\end{array}} \right])	
\end{eqnarray}
where $({x_{e1d}},{y_{e1d}})$ and $({x_{e2d}},{y_{e2d}})$ denote the target trajectories in the exoskeleton cartesian space. Substituting Eq. (7) into Eq. (16) and transforming it into angular space, we can simplify to obtain:
\begin{eqnarray} 
	\left[ \begin{array}{l} \nonumber
		{{\dot \theta }_{e1d}}\\
		{{\dot \theta }_{e2d}}
	\end{array} \right] = {\Gamma ^{ - 1}}\{ {\tau _{Jhd}} - [J_{hh}^T({K_{{{\rm hint}} }}\left[ \begin{array}{l}
		{x_{e1d}} - {x_{h1}}\\
		{y_{e1d}} - {y_{e1}}
	\end{array} \right] \\\nonumber
	- {D_{{{\rm hint}} }}\left[ \begin{array}{l}
		{{\dot x}_{h1}}\\
		{{\dot y}_{h1}}
	\end{array} \right])\\
	\quad \quad \quad \quad +J_{hk}^T({K_{k{\mathop{\rm int}} }}\left[ \begin{array}{l}
		{x_{e2d}} - {x_{h2}}\\
		{y_{e2d}} - {y_{h2}}
	\end{array} \right] + {D_{k{\mathop{\rm int}} }}\left[ \begin{array}{l}
		{{\dot x}_{h2}}\\
		{{\dot y}_{h2}}
	\end{array} \right])]\} 
\end{eqnarray}
where $\Gamma= J_{hk}^T{D_{k{\mathop{\rm int}} }}\left[ \begin{array}{l}
	{L_{e1d}}{C_{e1d}} + {l_{e2d}}{C_{e1de2d}}{\rm{  }}{l_{e2d}}{C_{e1de2d}}{\rm{ }}\\
	- {L_{e1d}}{S_{e1d}} - {l_{e2d}}{S_{e1e2}}{\rm{   }} - {l_{e2d}}{S_{e1e2}}
\end{array} \right]\\ + J_{hh}^T{D_{h{\mathop{\rm int}} }}\left[ \begin{array}{l}
\,\,\,{l_{e1d}}{C_{e1d}}\quad 0\\
- {l_{e1d}}{S_{e1d}}\quad 0
\end{array} \right] $, Therefore, it is only needed to integrate the matrix ${{\dot q}_{ed}} = {\left[ {{{\dot \theta }_{e1d}}\,{{\dot \theta }_{e2d}}} \right]^T}$ to obtain the target tracking trajectory $q_{ed}$ of the exoskeleton.

\subsection{APC Design}

Through the aforementioned analysis and discussion, the desired interaction torque has been effectively translated into the target trajectory of the exoskeleton. The subsequent steps will utilize the APC method to ensure accurate and rapid position tracking of the exoskeleton, thereby fulfilling the requirements for assisted rehabilitation. By combining Eqs. (1) and (7), the mathematical model of the exoskeleton is expressed as follows:
\begin{eqnarray}
	{M_e}({q_e}){{\ddot q}_e} + {C_e}({{\dot q}_e},{q_e}){{\dot q}_e} + {G_e}({q_e}) + {\tau _f}({{\dot q}_e}) = u
\end{eqnarray}
where $u = {\tau _{con}} + {\tau _{Je}}$, ${\tau _f}$ and ${\tau _{con}}$ denote joint friction and controller input torque, respectively. In accordance with literature [16], ${\tau _f}$ is specifically expressed as follows:
\begin{eqnarray}
{\tau _f} = ({f_{cj}} + ({f_{sj}} - {f_{cj}}){e^{ - |\frac{{{{\dot \theta }_{ej}}}}{{{\theta _{js}}}}{|^{\alpha}}}}){\mathop{\rm sgn}} ({{\dot \theta }_{ej}}) - {f_{vj}}{{\dot \theta }_{ej}}
\end{eqnarray}
where ${f_{cj}}$ denotes the coulomb friction parameter, ${f_{sj}}$ represents the static friction parameter, ${\theta_{ej}}$ is the relative sliding velocity, ${\theta_{js}}$ is the stribeck velocity, $\alpha$ is the exponential factor, and $f_{vj}$ is the coefficient of viscous friction.

Based on the above dynamic equations, consider analyzing the system from the energy perspective: the kinetic and potential energy expressions are as follows:
\begin{eqnarray}\nonumber
E &=& \frac{1}{2}\dot q_e^TM{{\dot q}_e} + {m_{e1}}g{l_{e1}}(1 - {S_{e1}}) \\
&+& {m_{e2}}g(1 - ({l_{e1}}{S_{e1}} + {l_{e2}}{S_{e1e2}}))
\end{eqnarray}
Inspired by Eq. (20), the following error function on energy is designed:
\begin{eqnarray}\nonumber
{V_E} &=& \frac{1}{2}\dot q_{err}^T{M_e}{{\dot q}_{err}} + {m_{e1}}g{l_{e1}}1 - {S_{e1}} \\
&+& {m_{e2}}g(1 - ({l_{e1}}{S_{e1}} + {l_{e1}}{S_{e1e2}}))
\end{eqnarray}
where ${{\dot q}_{err}} = {[{{\dot e}_{{\theta _1}}},{{\dot e}_{{\theta _2}}}]^T}$, ${{\dot e}_{{\theta _1}}} = {{\dot \theta }_{e1}} - {{\dot \theta }_{e1d}}$, ${{\dot e}_{{\theta _2}}} = {{\dot \theta }_{e2}} - {{\dot \theta }_{e2d}}$.
Taking the time derivative of the energy equation $E$, the expression is as follows:
\begin{eqnarray}\nonumber
{{\dot V}_E} &=& \dot q_{err}^T{M_e}{{\ddot q}_{err}} + \frac{1}{2}\dot q_{err}^T{{\dot M}_e}{{\dot q}_{err}}\\\nonumber
&-& ({m_{e1}}g{l_{e1}}{C_{e1}} + {m_{e2}}g{l_{e1}}{C_{e1}} + {m_{e2}}g{l_{e2}}{C_{e1e2}}){{\dot \theta }_{e1}}\\
&-& {m_{e2}}g{l_{e2}}{C_{e1e2}}{{\dot \theta }_{e2}}
\end{eqnarray}
Substituting Eq. (18) into Eq. (22) and combining with ${{\dot M}_e}({{\dot q}_e}) = {C_e}({{\dot q}_e},{q_e}) + C_e^T({{\dot q}_e},{q_e})$, the expression is:
\begin{eqnarray}\nonumber
{{\dot V}_E} &=& \dot q_{err}^T(u - {C_e}{{\dot q}_e} - {G_e} - {\tau _f} - M{{\ddot q}_d} + {C_e}({{\dot q}_e} - {{\dot q}_{ed}}))\\\nonumber
{\rm{   }} &-& ({m_{e1}}g{l_{e1}}{C_{e1}} + {m_{e2}}g{l_{e1}}{C_{e1}} + {m_{e2}}g{l_{e2}}{C_{e1e2}}){{\dot \theta }_{e1}} \\\nonumber
&-& {m_{e2}}g{l_{e2}}{C_{e1e2}}{{\dot \theta }_{e2}}\\\nonumber
{\rm{    }} &=& \dot q_{err}^T(u - {\tau _f} - {M_e}{{\ddot q}_{ed}} - {C_e}{{\dot q}_{ed}} - {G_e})\\\nonumber
{\rm{    }} &-& ({m_{e1}}g{l_{e1}}{C_{e1}} + {m_{e2}}g{l_{e1}}{C_{e1}} + {m_{e2}}g{l_{e2}}{C_{e1e2}}){{\dot \theta }_{e1}}\\\nonumber
 &-& {m_{e2}}g{l_{e2}}{C_{e1e2}}{{\dot \theta }_{e2}}\\\nonumber
{\rm{    }} &=& \dot q_{err}^T( u - {\tau _f} - {M_e}{{\ddot q}_{ed}} - {C_e}{{\dot q}_{ed}}) \\
&+& (\dot q_e^T - \dot q_{ed}^T)\left[ \begin{array}{l}
{{G_{e1}}}\\\nonumber
	{{G_{e2}}}
\end{array} \right] + \dot q_e^T\left[ \begin{array}{l}
	{{G_{e1}}}\\
{{G_{e2}}}
\end{array} \right]\\
{\rm{    }} &=& \dot q_{err}^T(u - {\tau _f} - {M_e}{{\ddot q}_{ed}} - {C_e}{{\dot q}_{ed}}) + G_e{{\dot q}_{ed}}
\end{eqnarray}
where ${M_e}{{\ddot q}_{ed}}$, ${C_e}{{\dot q}_{ed}}$ and ${\tau _f}$ are given below:
\begin{eqnarray}\nonumber
&{M_e}{{\ddot q}_{ed}} = \left[ \begin{array}{l}
	{n_1} + {n_2} + 2{n_3}{C_{e2}}\quad{n_2} + {n_3}{C_{e2}}\\
	{n_2} + {n_3}{C_{e2}}\quad\quad{n_2}
\end{array} \right]\left[ \begin{array}{l}
	{{\ddot \theta }_{e1d}}\\
	{{\ddot \theta }_{e2d}}
\end{array} \right]\\\nonumber
&{C_e}{{\dot q}_{ed}} = \left[ \begin{array}{l}
	- 2{n_3}{S_{e2}}{{\dot \theta }_{e2}}\quad\quad - {n_3}{S_{e2}}{{\dot \theta }_{e2}}\\
	{n_3}{S_{e2}}{{\dot \theta }_{e1}}\quad\quad\quad\quad0
\end{array} \right]\left[ \begin{array}{l}
	{{\dot \theta }_{e1d}}\\
	{{\dot \theta }_{e2d}}
\end{array} \right]\\\nonumber
&{\tau _f} = \left[ \begin{array}{l}
	{n_4}{\mathop{\rm sgn}} ({{\dot \theta }_{e1}}) - {n_5}{{\dot \theta }_{e1}}\\
	{n_6}{\mathop{\rm sgn}} ({{\dot \theta }_{e2}}) - {n_7}{{\dot \theta }_{e2}}
\end{array} \right]
\end{eqnarray}
and ${n_1} = {m_{e1}}l_{e1}^2 + {m_{e2}}L_{e1}^2 + {J_{e1}}$; ${n_2} = {m_{e2}}l_{e2}^2 + {J_{e2}}$; ${n_3} = {m_{e2}}{L_{e1}}{l_{e2}}$, ${n_4} = {f_{c1}} + ({f_{s2}} - {f_{c2}}){e^{ - |\frac{{{{\dot \theta }_{2d}}}}{{{\theta _{2s}}}}{|^{\alpha}}}}$; ${n_5} = {f_{v1}}$; ${n_6} = {f_{c2}} + ({f_{s2}} - {f_{c2}}){e^{ - |\frac{{{{\dot \theta }_{2d}}}}{{{\theta _{2s}}}}{|^{\alpha}}}}$; ${n_7} = {f_{v2}}$.

Transform ${M_e}{{\ddot q}_{ed}} + {C_e}{{\dot q}_{ed}} + {\tau _f}$ in Eq. (23) into the following form:
\begin{eqnarray}
{M_e}{{\ddot q}_{ed}} + {C_e}{{\dot q}_{ed}} + {\tau _f} = \left[ \begin{array}{l}
	\psi _h^T{\omega _h}\\
	\psi _k^T{\omega _k}
\end{array} \right]
\end{eqnarray}
where ${\psi _h} = {\left[ {\,{\psi _{h1}}\,{\psi _{h2}}\,{\psi _{h3}}\,{\psi _{h4}}\,{\psi _{h5}}} \right]^T}$ and ${\psi _k} = {\left[ {{\psi _{k1}}\,{\psi _{k2}}\,{\psi _{k3}}\,{\psi _{k4}}} \right]^T}$ denote the adaptive separation terms, and ${\omega _h} = {\left[ {{n_1}\,{n_2}\,{n_3}\,{n_4}\,{n_5}} \right]^T}$ and ${\omega _k} = {\left[ {{n_2}\,{n_3}\,{n_6}\,{n_7}} \right]^T}$ represent time-varying parameters of the estimated system during motion. The specific separation expression is as follows:
\begin{eqnarray}\nonumber
&{\psi _{h1}} = {{\ddot \theta }_{e1d}};{\psi _{h2}} = {{\ddot \theta }_{e1d}} + {{\ddot \theta }_{e2d}};\\\nonumber
&{\psi _{h3}} = 2{C_{e2}}{{\ddot \theta }_{e1d}} + {C_{e2}}{{\ddot \theta }_{e2d}} - 2{S_{e2}}{{\dot \theta }_{e2}}{{\dot \theta }_{e1d}} - {S_{e2}}{{\dot \theta }_{e2}}{{\dot \theta }_{e2d}};\\
&{\psi _{h4}} =  - {\mathop{\rm sgn}} ({{\dot \theta }_{e1}});{\psi _{h5}} =  - {{\dot \theta }_{e1}}\\\nonumber
&{\psi _{k1}} = {{\ddot \theta }_{1ed}} + {{\ddot \theta }_{e2d}}{\psi _{k2}} = {C_{e2}}{{\ddot \theta }_{e1d}} + {S_{e2}}{{\dot \theta }_{e1}}{{\dot \theta }_{e1d}}\\
&{\psi _{k3}} =  - {\mathop{\rm sgn}} ({{\dot \theta }_{e2}}){\psi _{k4}} =  - {{\dot \theta }_{e2}}
\end{eqnarray}
Bringing Eq. (24) into (23) simplifies to:
\begin{eqnarray}
{{\dot V}_E} = ({u_1} - \psi _h^T{\omega _h}){{\dot e}_{{\theta _1}}} + ({u_2} - \psi _k^T{\omega _k}){{\dot e}_{{\theta _2}}} + G{{\dot q}_{ed}}
\end{eqnarray}
Due to the continuous variation of system parameters during the exoskeleton's motion and the limited accuracy of parameter identification, we estimate these parameters by introducing weighting factors $w_1$ and $w_2$. Based on this, the following nonlinear controller is designed:
\begin{eqnarray}
{u_1} &=&  - {k_{ph}}{e_{{\theta _1}}} - {k_{dh}}{{\dot e}_{{\theta _1}}} + \psi _h^T{{\hat \omega }_h} - G_{e1}\\
{u_2} &=&  - {k_{pk}}{e_{{\theta _2}}} - {k_{dk}}{{\dot e}_{{\theta _2}}} + \psi _k^T{{\hat \omega }_k} - G_{e2}
\end{eqnarray}
where ${k_{ph}},{k_{dh}},{k_{pk}},{k_{dk}} \in {R^ + }$ represent positive gain control parameters. ${{\hat \omega }_h}$ and ${{\hat \omega }_k}$ represent the estimated values of the uncertain vectors ${\omega _h}$ and ${\omega _k}$, and their mathematical expressions are as follows:
\begin{eqnarray}
{{\hat \omega }_h} &=& {\left[ {{{\hat n}_1}\quad{{\hat n}_2}\quad{{\hat n}_3}\quad{{\hat n}_4}\quad{{\hat n}_5}} \right]^T}\\
{{\hat \omega }_k} &=& {\left[ {{{\hat n}_2}\quad{{\hat n}_3}\quad{{\hat n}_6}\quad{{\hat n}_7}} \right]^T}
\end{eqnarray}
The parameter update rate is as follows:
\begin{eqnarray}
{{\dot {\hat \omega} }_h} =  - \Pi _h^{ - 1}{\psi _h}{{\dot e}_{{\theta _1}}}\\
{{\dot {\hat \omega }}_k} =  - \Pi _k^{ - 1}{\psi _k}{{\dot e}_{{\theta _2}}}
\end{eqnarray}
where ${\Pi _h} = diag\{ {\Pi _{h1}},...,{\Pi _{h5}}\}  \in {R^ + }$ and ${\Pi _k} = diag\{ {\Pi _{k1}},...,{\Pi _{k4}}\}  \in {R^ + }$, and their parameters can be designed.

\section{Simulation and discussion }
\subsection{Proposed controller}
A mathematical model is built in Matlab/simulink to verify the effectiveness of the proposed control method. Where the target recovery trajectory is derived from literature [5] and optimised as follows:
\begin{eqnarray}
{\theta _{h1}} = 90 - \mathop \sum \limits_{i = 1}^4 {a_h}[{\rm I}]\sin ({\rm I}\omega t) - \mathop \sum \limits_{i = 1}^4 {b_h}[{\rm I}]\cos ({\rm I}\omega t) + {q_{h0}}\\
{\theta _{h2}} = \mathop \sum \limits_{i = 1}^4 {a_k}[{\rm I}]\sin ({\rm I}\omega t) + \mathop \sum \limits_{i = 1}^4 {b_k}[{\rm I}]\cos ({\rm I}\omega t) + {q_{k0}}
\end{eqnarray}
where $\omega  = 0.4\pi$, ${a_{\varepsilon ,hip}}(\varepsilon  = 1,...,4)$ are -2.874, -2.423, 1.227, -0.1462, respectively, ${b_{\varepsilon ,hip}}(\varepsilon  = 1,...,4)$ are 18.53, -2.016,-0.3704, 0.201 respectively, ${a_{\varepsilon ,knee}}(\varepsilon  = 1,...,4)$ are 17.62,-2.469,-3.82, -0.1346 respectively, ${b_{\varepsilon ,knee}}(\varepsilon  = 1,...,4)$ are -1.494,11.72,  1.014, 0.2165 respectively, ${q_{h0}}=10.07$ and ${q_{k0}}=-17.49$. The parameters of
$u$ are set as $k_{ph}$ = 500, $k_{dh}$ = 50, $k_{pk}$ = 100, $k_{dk}$ = 25, ${\Pi _h} = diag[10000\, 10000\, 1000\, 100\, 1000]$,  ${\Pi _k} = diag[10\,10\, 10\, 10]$. The parameters of
$\tau_f$ are set as $f_{cj}$ = 4.27, $f_{sj}$ = 11.55, $\theta_{js}$ = 107.17, $f_{vj}$ = 0.067,  $\alpha$ = -1.38.The specific system parameters are shown in \textbf{Table 2}.

\begin{table}[t]
	\begin{center}
		\renewcommand{\arraystretch}{1.2} 
		\caption{System model parameters values}
		\label{tab_1}
		\begin{tabular}{c c c c c c}
			\hline
			\scriptsize ${L_{h1}}$ [m] & \scriptsize 0.38 & \scriptsize ${L_{e1}}$ [m] & \scriptsize 0.38 & \scriptsize ${l_{h1}}$ [m] & \scriptsize 0.25 \\
			\hline
			\scriptsize ${l_{e1}}$ [m] & \scriptsize 0.25 & \scriptsize $m_{h1}$ [kg] & \scriptsize 8.171 & \scriptsize $m_{e1}$ [kg] & \scriptsize 1.173 \\
			\scriptsize ${J_{h1}}$ [$\mathrm{kgm^2}$] & \scriptsize 0.123 & \scriptsize $J_{e1}$ [$\mathrm{kgm^2}$] & \scriptsize 0.016 & \scriptsize $L_{h2}$ [m] & \scriptsize 0.46 \\
			\scriptsize ${L_{e2}}$ [m] & \scriptsize 0.39 & \scriptsize $l_{h2}$ [m] & \scriptsize 0.2 & \scriptsize $l_{e2}$ [m] & \scriptsize 0.2 \\
			\scriptsize $m_{h2}$ [kg] & \scriptsize 3.859 & \scriptsize $m_{e2}$ [kg] & \scriptsize 0.853 & \scriptsize $J_{h2}$ [$\mathrm{kgm^2}$] & \scriptsize 0.109 \\
			\scriptsize $J_{e2}$ [$\mathrm{kgm^2}$] & \scriptsize 0.008 & \scriptsize $g$ [m/s$^2$] & \scriptsize 9.8 & \scriptsize $K$ [N/m] & \scriptsize 500 \\
			\scriptsize $D$ [Ns/m] & \scriptsize 1.25 & \scriptsize - & \scriptsize - & \scriptsize - & \scriptsize - \\
			\hline
		\end{tabular}
	\end{center}
\end{table}

\begin{figure}[!htb]
	\centering
	\includegraphics[width=\hsize]{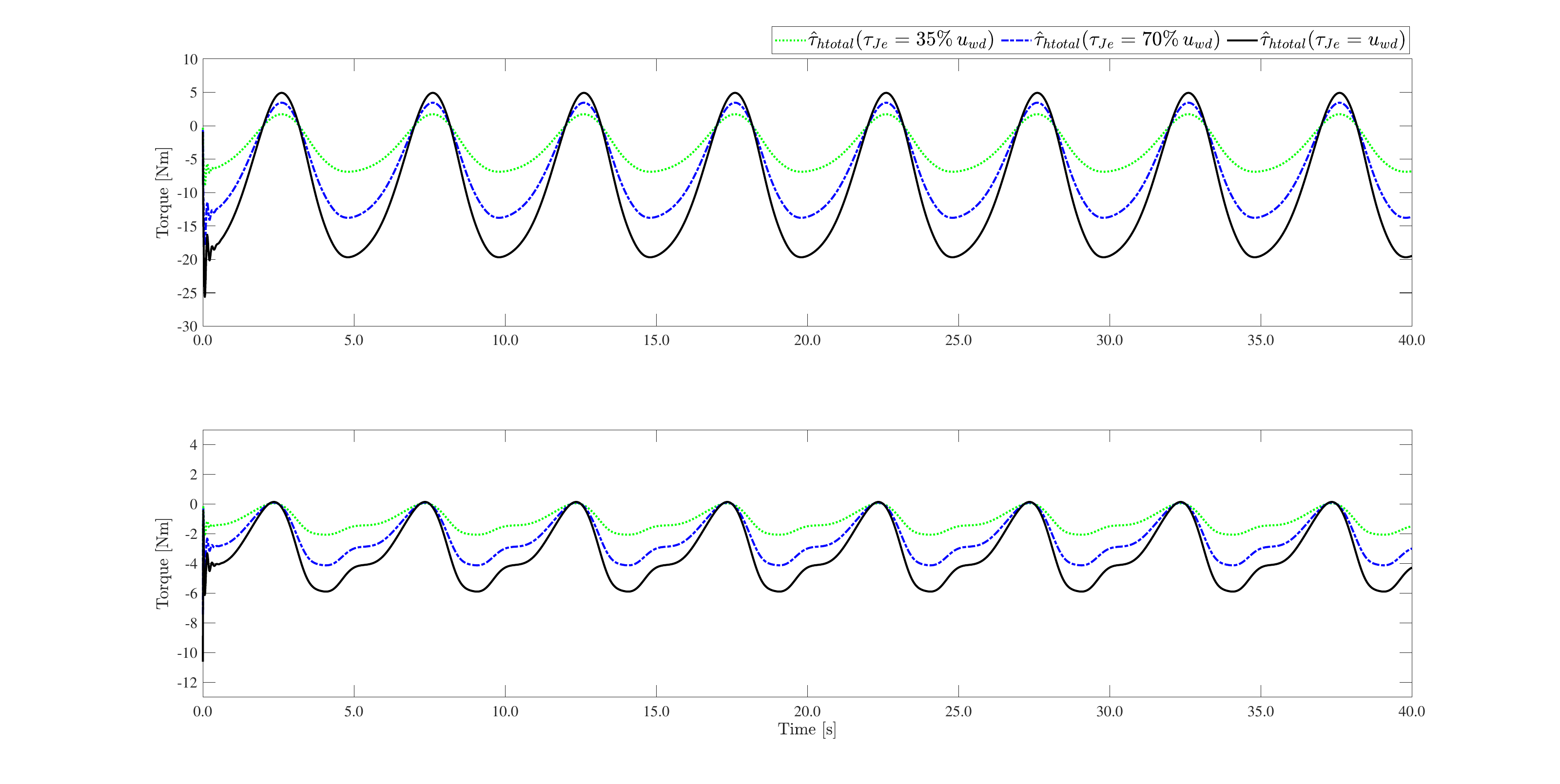}
	\caption{HTMO when the human muscle moment is 35\%, 70\%, 100\% of the target torque. upper: hip, lower: knee.}
	\label{figure3}
\end{figure}


\begin{figure}[!htb]
	\centering
	\includegraphics[width=\hsize]{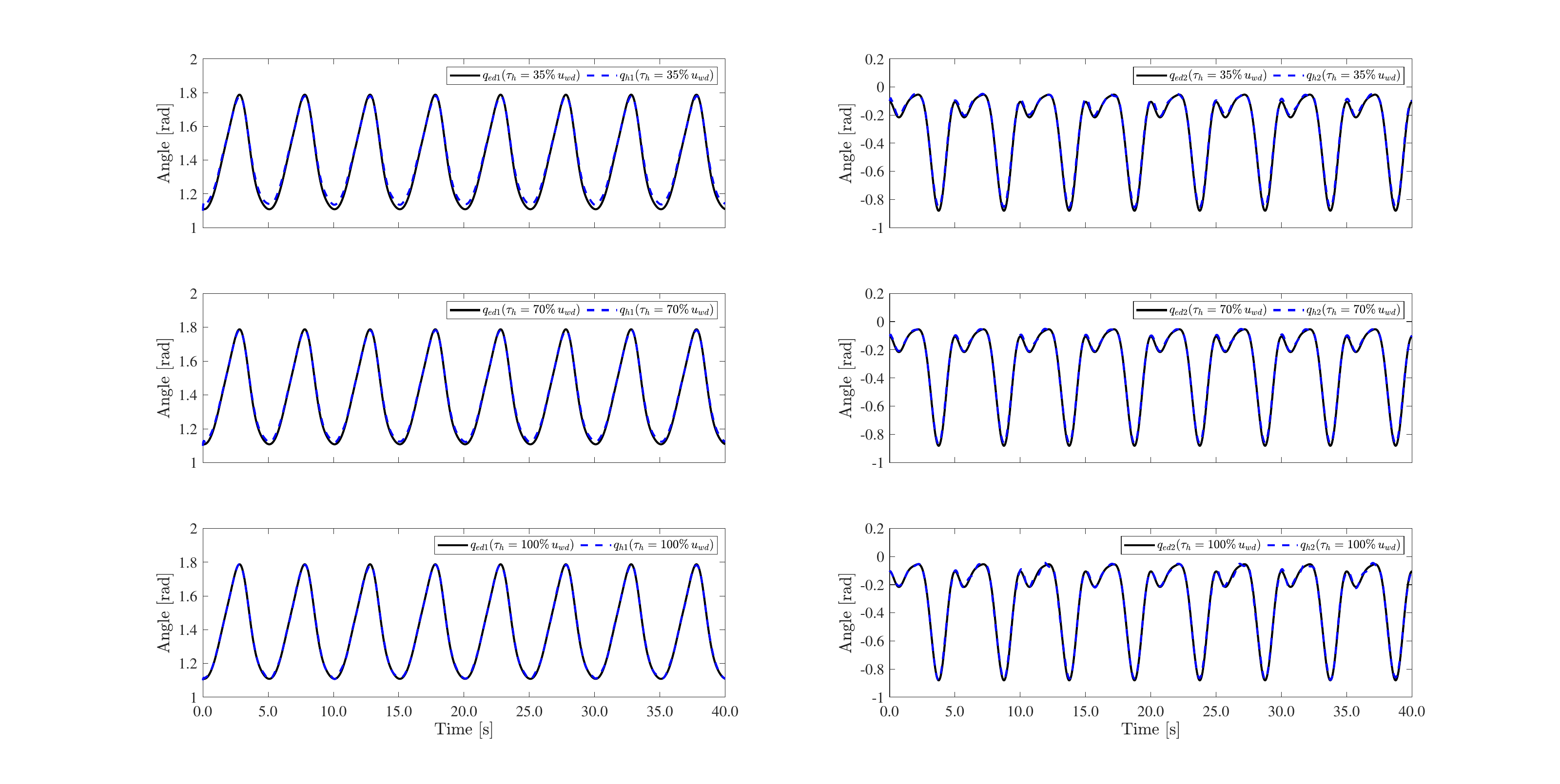}
	\caption{Trajectory tracking chart of the wearer under different torques provided by the patient. left: hip, right: knee.}
	\label{figure4}
\end{figure}

To analyze the effectiveness of the APC when the wearer's lower limb mobility changes, ${\tau_h}$ is set to 35\%, 70\%, and 100\% of the target input torque, simulating different levels of lower limb mobility of the wearer. ${\tau_h}$  = 35\%, 70\%${\tau_{Jhd}}$. ${\tau_h}$ indicates that the wearer has varying degrees of lower limb muscle weakness.  ${\tau_h}$ = 100\% ${\tau_{Jhd}}$ indicates that the wearer has good lower limb health.

In Fig. 3, the estimated ${{\hat \tau }_{htotal}}$ of the HTMO is shown under different levels of the wearer’s motion capabilities. According to the paper ${\tau_{htotald}}$ consists of ${\tau_{htotal}}$ and ${\tau_{Jhd}}$, and from Fig. 3, it can be seen that ${\hat \tau_{htotal}}$ approximates ${\tau_h}$. The HMMO can qualitatively reflect the torque exerted by the patient. Then, ${\tau_{htotald}}$ is subtracted from ${\hat \tau_{htotal}}$, which can visualise the moment that the exoskeleton needs to provide. Additionally, the closer ${\hat \tau_{htotal}}$ is to ${ \tau_{htotald}}$, the smaller the interactive force the exoskeleton needs to provide, indicating a stronger motion capability of the patient. Based on this, from Fig. 4, it can be seen that when the torque provided by the human body is 100\% $\tau_{htotald}$, the errors in $q_{ed}$ and $q_h$ should be minimized. This also validates the relationship between the interaction torque and the target torque. It is worth mentioning that the exoskeleton trajectory $q_{ed}$ is obtained from Eq. (17).

\subsection{Contrast Controllers}
In this section, the proposed controller is compared and analyzed with existing controllers (LQR, AANSMC). First, the linearized model is presented to prepare for the subsequent controller design:
\begin{eqnarray}
M_e^*{{\ddot q}_e} + G_e^* = u	
\end{eqnarray}
where $M_e^* = \left[ {m_{ij}^*} \right] \in {R^{2 \times 2}}$. $i = 1,2,j = 1,2$. and $M_{e11}^* = {m_2}L_1^2 + 2{m_2}{L_1}{l_2} + {m_1}l_1^2 + {m_2}l_2^2 + {J_1} + {J_2}$; $M_{e12}^* = M_{e21}^* = {m_2}l_2^2 + {L_1}{m_2}{l_2} + {J_2}$; $M_{e22}^* = {m_2}l_2^2 + {J_2}$. $G_e^* = {\left[ {G_{e1}^*G_{e2}^*} \right]^T}$, where $G_{e1}^*{\rm{ }} =  - g{l_2}{m_2}{\rm{ }} - {\rm{ }}{L_1}g{m_2}{\rm{ }} - {\rm{ }}g{l_1}{m_1}$; $G_{e2}^*{\rm{ }} =  - {\rm{ }}g{l_2}{m_2}$.

Then, the LQR[14] controller is designed as follows:
\begin{eqnarray}
u_1^{lqr} =  - k_{11}^l{e_{{\theta _1}}} - k_{12}^l{e_{{\theta _2}}} - k_{13}^l{{\dot e}_{{\theta _1}}} - k_{14}^l{{\dot e}_{{\theta _2}}} + \tau_{f1}\\
u_2^{lqr} =  - k_{21}^l{e_{{\theta _1}}} - k_{22}^l{e_{{\theta _2}}} - k_{23}^l{{\dot e}_{{\theta _1}}} - k_{24}^l{{\dot e}_{{\theta _2}}}+ \tau_{f2}
\end{eqnarray}
where $\tau_{f1}$ and $\tau_{f2}$ are friction compensation. By analysing the system, the parameters $Q$ = diag[2000\, 2000 \,2\, 2] and $R$ = diag[0.1\,0.1] were selected to obtain the control parameters as follows: $
k_{11}^l = 5458.8$, $k_{12}^l = 1.6$, $k_{13}^l = 0.51$, $k_{14}^l = 0.28$,	$k_{21}^l = 0.15$, $k_{22}^l = 1045.3$, $k_{23}^l =  - 0.3$, $k_{24}^l = 0.75$.

\begin{figure}[!htb]
	\centering
	\includegraphics[width=\hsize]{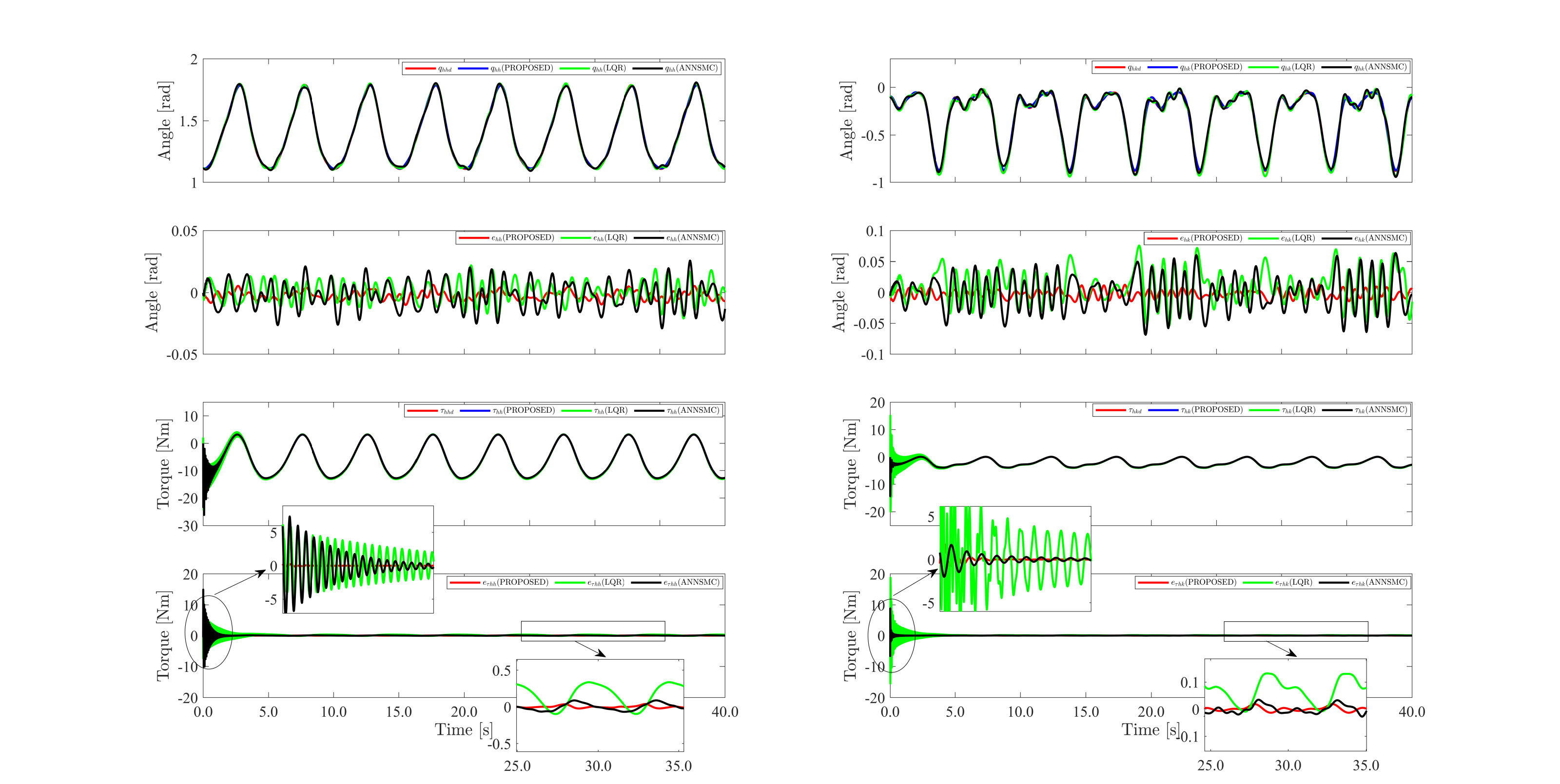}
	\caption{Trajectory/interaction torque tracking curves under different controllers with 35\% of patients providing torque. left: hip, right: knee.}
	\label{figure6}
\end{figure}

\begin{figure}[!htb]
	\centering
	\includegraphics[width=\hsize]{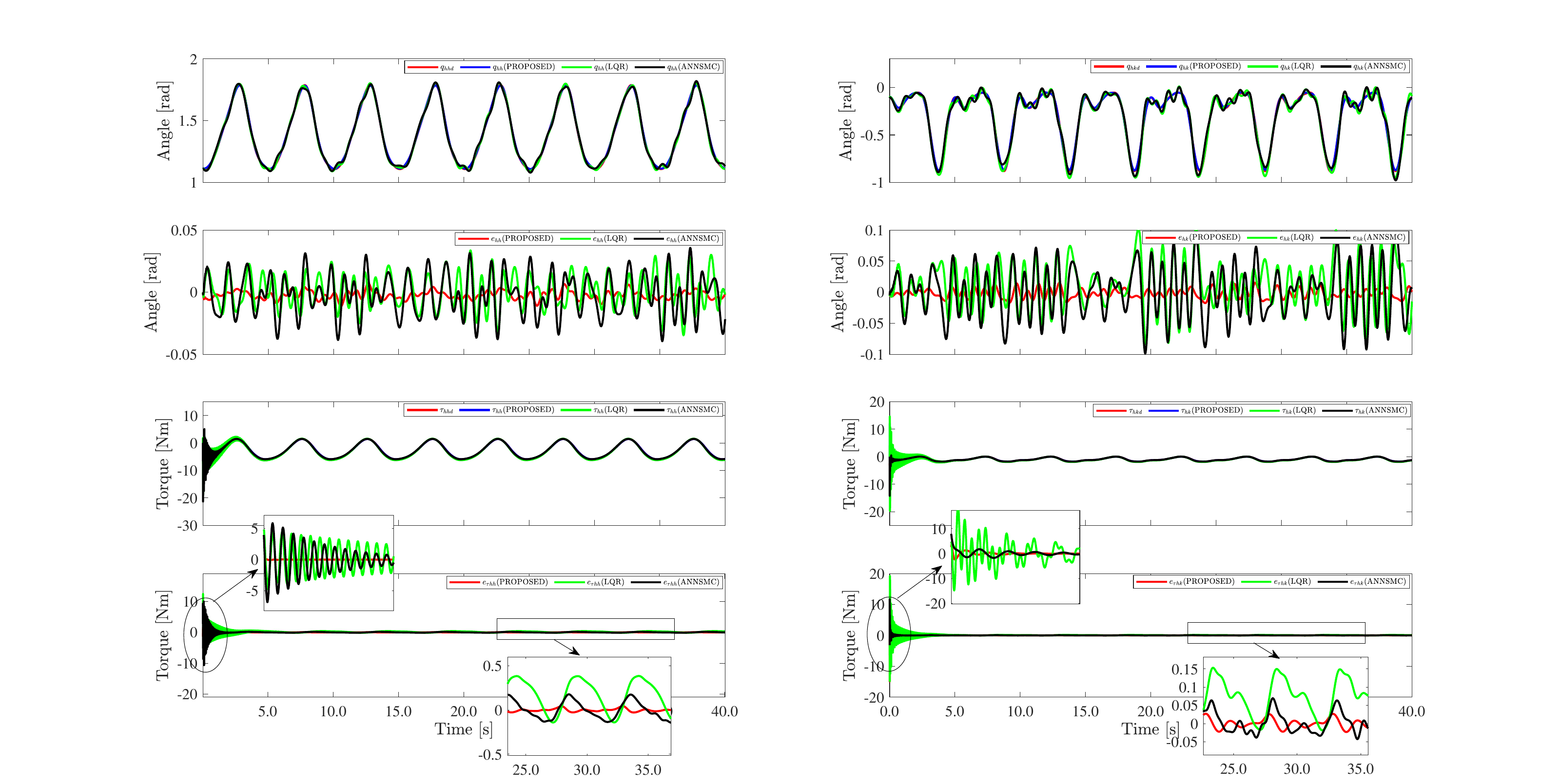}
	\caption{Trajectory/interaction torque tracking curves under different controllers with 70\% of patients providing torque. left: hip, right: knee.}
	\label{figure7}
\end{figure}

\begin{figure}[!htb]
	\centering
	\includegraphics[width=\hsize]{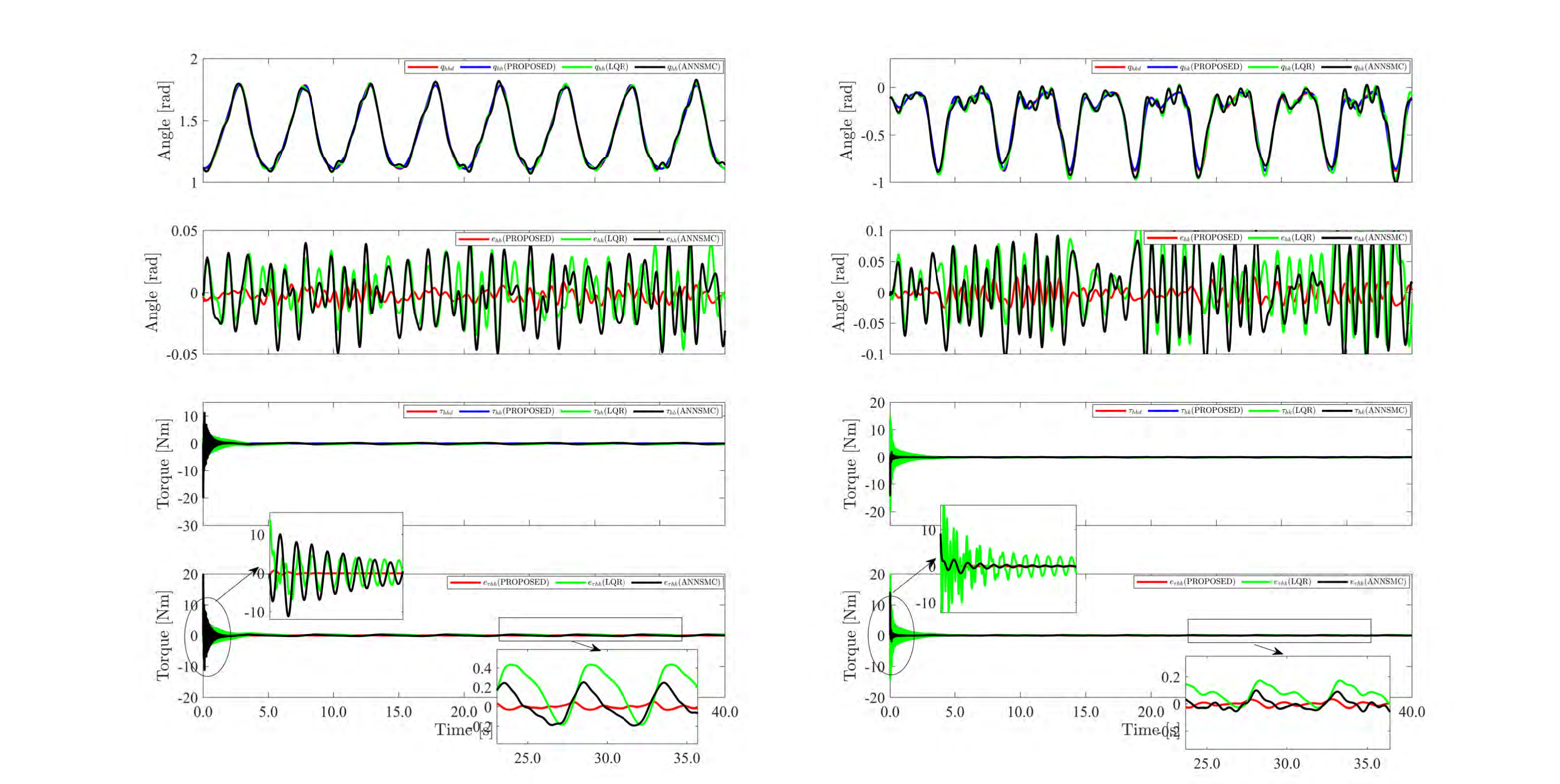}
	\caption{Trajectory/interaction torque tracking curves under different controllers with 100\% of patients providing torque. left: hip, right: knee.}
	\label{figure8}
\end{figure}

The design of the second controller(AANSMC)[8] is as follows:
\begin{eqnarray}
u{\rm{ }} = {\rm{ }}{M_e}\left( {{q_{ed}} - {K_v}{{\dot e}_\theta } - {K_p}{e_\theta } - \rho {\mathop{\rm sgn}} {{\dot e}_\theta }} \right) + {C_e}{{\dot q}_e} + {\rm{ }}G_e
\end{eqnarray}
where $K_p$=diag[850\,660], $K_v$=diag[5\,20], $\rho$=diag[0.5\,0.5]. 

From Fig. (5)-(7), the tracking errors between the rehabilitation motion trajectory and interaction torque under different torques provided by the patients are compared. A comparative analysis with LQR and AANSMC controls was conducted. The proposed control shows good performance in both the hip and knee joints. In the trajectory tracking of rehabilitation patients, the torque provided by the patient ranges from 35\% to 100\%, and the proposed controller's trajectory tracking error can always be controlled within 0.02 rad.  Compared with other controllers, LQR has an average of approximately 0.055 rad, while AANSMC is 0.5 rad. Additionally, in the interaction torque tracking system, whether in the initial tracking state (see the enlarged image) or the final steady state, compared to the severe oscillations of LQR and AANSMC, the proposed controller's torque output remains relatively smooth, ensuring the performance of the controller and the safety of the patient.

\section{Conclusion}

In this paper, we have proposed an innovative AAN control strategy for lower limb exoskeletons, aimed at enhancing rehabilitation outcomes. The approach integrates a human-machine coupling dynamics model, a HTMO, and APC, ensuring a dynamic and user-specific rehabilitation process. By effectively estimating joint torques and providing adaptive control, the system adjusts to varying patient conditions, offering personalized assistance. Simulation results demonstrate that the proposed strategy outperforms traditional controllers, including LQR and AANSMC, in terms of trajectory tracking accuracy and interaction torque smoothness. This approach not only minimizes tracking errors but also improves the safety and efficiency of rehabilitation, making it a promising solution for enhancing the therapeutic potential of exoskeleton-assisted rehabilitation.

\end{document}